\documentclass[final,5p,times,twocolumn]{elsarticle}

\usepackage{lineno}
\usepackage{graphicx}
\usepackage{amssymb}
\usepackage{pdflscape}
\usepackage[english]{babel}
\usepackage[dvipsnames,svgnames,x11names]{xcolor}
\usepackage{booktabs,tabularx}
\usepackage{multirow}
\usepackage{subfig} 
\usepackage{hyperref}
\usepackage{amsmath}
\usepackage{commath}
\usepackage[english]{babel}
\usepackage[ruled]{algorithm2e}
\SetEndCharOfAlgoLine{}
\usepackage[leftmargin=6em,rightmargin=12em,indentfirst=false]{quoting}
\usepackage{siunitx}
\usepackage{comment} 

\usepackage[final]{changes}
\usepackage{float}

\usepackage{lineno}
\usepackage{tikz}
\usepackage{comment} 
\usepackage[export]{adjustbox}

\usepackage{graphicx}
\usepackage[utf8]{inputenc}
\usepackage[export]{adjustbox}
\usepackage{wrapfig}
\usepackage{dcolumn}

\makeatletter
\newcolumntype{T}[3]{>{\textfont0=\the@{#1}{#2}{#3}}c<{\DC@end}}
\makeatother

\usepackage{pgfplots}
\pgfplotsset{width=10cm,compat=1.9}

\usepackage{array}

\newcolumntype{L}[1]{>{\raggedright\let\newline\\\arraybackslash\hspace{0pt}}m{#1}}
\newcolumntype{C}[1]{>{\centering\let\newline\\\arraybackslash\hspace{0pt}}m{#1}}
\newcolumntype{R}[1]{>{\raggedleft\let\newline\\\arraybackslash\hspace{0pt}}m{#1}}

\usepackage{subfiles}

\usepackage{todonotes}

\journal{Frontiers in Built Environment}
\begin{document}
	
\begin{frontmatter}

\title{Spacematch: Using environmental preferences to match occupants to suitable activity-based workspaces}

\author{Tapeesh Sood\,$^{1}$, Patrick Janssen\,$^{2}$ and Clayton Miller\,$^{1,*}$}

\address{$^{1}$Building and Urban Data Science (BUDS) Lab, Department of Building, School of Design and Environment (SDE), National University of Singapore (NUS), Singapore}
\address{$^{2}$Department of Architecture, School of Design and Environment (SDE), National University of Singapore (NUS), Singapore}
\address{$^*$Corresponding Author: clayton@nus.edus.sg, +65 81602452}

\begin{abstract}
The activity-based workspace (ABW) paradigm is becoming more popular in commercial office spaces. In this strategy, occupants are given a choice of spaces to do their work and personal activities on a day-to-day basis. This paper shows the implementation and testing of the \emph{Spacematch} platform that was designed to improve the allocation and management of ABW. An experiment was implemented to test the ability to characterize the preferences of occupants to match them with suitable environmentally-comfortable and spatially-efficient flexible workspaces. This approach connects occupants with a catalog of available work desks using a web-based mobile application and enables them to provide real-time environmental feedback. In this work, we tested the ability for this feedback data to be merged with indoor environmental values from Internet-of-Things (IoT) sensors to optimize space and energy use by grouping occupants with similar preferences. This paper outlines a case study implementation of this platform on two office buildings. This deployment collected 1,182 responses from 25 field-based research participants over a 30-day study. From this initial data set, the results show that the ABW occupants can be segmented into specific \emph{types} of users based on their accumulated preference data, and matching preferences can be derived to build a recommendation platform.
\end{abstract}

\end{frontmatter}


\section{Introduction}
In the past few years, rising corporate real estate (CRE) costs and rapid changes in technology and nature of work have rendered inefficient traditional modes of working \emph{where occupants are permanently designated a single work desk}. Today, 37\% of all office spaces are empty on any given workday \citep{jll}, which equates to approximately 150 billion dollars annually in unused space globally \citep{cbre}. These challenges are pushing building operators to rethink occupant density and spatial utilization in workplaces. In a recent survey, while 95\% of CRE professionals believed that workplaces influence occupant productivity and comfort, only one third measured that impact. Most others only considered traditional cost-based measures and metrics to quantify workplace density and utilization \citep{gensler}. Based on past studies, heavy reliance on cost-based measures generally results in operators taking away amenities that make occupants comfortable and productive in cost-saving work environments. An example is the replacement of cubicles with benches to accommodate a more significant headcount, removal of informal collaboration spaces for more desks, or taking away of employee storage space all together \citep{cbre}. Workplaces today face challenges as operators look to offset high rental costs by growing occupant headcount within their existing footprint. Equally, rapid changes in technology and the nature of work in the past few years have led them to rethink spatial density and utilization.

\subsection{The emergence of workplace flexibility}
In response to these challenges, new ways of working are evolving rapidly. These approaches aspire to simultaneously balance operator's \emph{cost and space saving} demands with flexibility and comfort needs of employees through enabling occupant mobility. Through most of these approaches, the occupant can work flexibly by choosing different spots within the workplace rather than being assigned a fixed desk as the one primary place of work. Once occupants are dynamic in the ways they use space, it is easier to recapture underutilized spaces by operators \citep{cbre}. Workplace strategies of this kind are often referred to as \emph{activity-based workspaces (ABW)} or by other terms such as hot-desking, co-working, desk-sharing, flexible working, and office hoteling. Though each strategy varies slightly from the other, most promise benefits of improved spatial utilization and cost savings for operators while increasing overall comfort, choice, and control for occupants \citep{Engelen2018IsReview}. Equally, an extensive recent survey of spaces utilizing one such strategy showed that the primary motivations for occupants to work in such a workplace is because it allowed access to an inspiring work environment \citep{Weijs-Perree2018AnalysingCharacteristics}. Understandably, an increase in the adoption of these concepts can be seen in the growing co-working industry.

\subsubsection{COVID-19 and its effect on the future of work}
In early 2020, office working habits of much of the world changed due to a global pandemic of a novel pathogen known as \emph{COVID-19}. As there is no treatment or vaccine for this virus, numerous community-driven mitigation strategies have been deployed across the world. One of the most common is the requirement for those that can work from home to do so \citep{Ebrahim2020-he}. This exodus from office spaces to the home has shown that such decentralization of office work is possible, and desirable in some situations. Going forward this forced push towards home working will reinforce the need for corporate entities to adopt more agile real-estate portfolios, with more flexibility and a more distributed footprint to ease employees back into the workplace and cut commuting time to a single, large headquarters \citep{jll_strategy_2020}. This digitization and rethinking of how people can work could be a strong catalyst for the adoption of ABW-style office arrangements.

\subsubsection{The limitations of ABW strategies}
Despite the momentum towards ABW, there are a significant number of challenges that this strategy poses through its shift in office culture. A study focused on a sociological analysis of one approach showed a "loss of everyday workspace ownership giving rise to practical and social tensions within the organization \citep{Hirst2011SettlersHot-desking}". Another study found lower than expected satisfaction with activity-based working environments due to rare switching of different activity settings \citep{Hoendervanger2016FlexibilityEnvironments}. A recent study even found evidence of \emph{dehumanization} as a result of ABW \citep{Taskin2019-ee}. As organizations evolve, they are wary of ill-conceived applications \emph{which may disrupt business and culture purely for cost savings} of new workplace strategies. These studies illustrate that there is much improvement possible in the deployment of ABW strategies to help mitigate these downsides.

\subsection{Connection to indoor environmental comfort in buildings}
In addition to space use allocation, indoor environmental comfort is at the forefront of building performance analysis. Occupant dissatisfaction with indoor environments has far-reaching economic implications for workplaces. As people typically spend over 90\% of their time indoors, indoor environment quality influences their comfort, performance, health, and well-being. Occupant dissatisfaction with indoor environments can result in health impacts, absenteeism, and reduced productivity \citep{MiltonDonaldK.P.MarkGlencross2000}. Not only do enterprises today associate a majority of their costs (80-90\%) to workers compensation and benefits \citep{Creativeandproductiveworkplaces, kats2003green, wilson2005making} but as people typically spend more (about 90\%) time indoors, the quality of the indoor environments influences their comfort, performance and well-being at work. Continual occupant dissatisfaction with indoor environments can result in health impacts, absenteeism, and reduced productivity \citep{MiltonDonaldK.P.MarkGlencross2000}. A comparison of recent field studies from 467 air-conditioned buildings containing 24,000 occupants showed between 30\% and 200\% more cases of sick building syndrome symptoms than in the occupants of naturally ventilated buildings \citep{Evolvingopportunities, ventilationsystemtype}. Another survey in 2012, of 52,980 occupants in 351 office buildings, found that 50\% of the occupants were dissatisfied with their indoor environments \citep{Frontczak2012QuantitativeDesign}.

In response to these challenges, research in indoor occupant comfort has accelerated over the last twenty years. One of the many paradigm shifts has been the movement away from traditional, physically-based deterministic models due to their reported low accuracy (only 34\%) across dozens of comfort studies in the past decades  \citep{CHEUNG2019205,FOLDVARYLICINA2018502}. Recent research efforts have progressed towards adaptive comfort models \citep{Ferrari2012AdaptiveIndices, Nicol2013AdaptiveWorld, vanHoof2010ThermalPractice}, which rely on human behavior. According to these models, discomforting changes in the thermal environment are followed by a behavioral change in people to restore comfort. Such actions could include reducing individual activity levels or even opening a window. The main effect of such models is to increase the range of conditions that designers can consider comfortable, for instance, naturally ventilated buildings in the tropics where occupants have a higher degree of control over their thermal environment.

Despite these advancements, even adaptive comfort models follow a one-size-fits-all approach that ignores the personal aspect of comfort; this is like expecting everyone to have similar preferences for food, music, or style; all subjective attributes of a person's personality. Further work on adaptive models is needed to identify comfort preferences on an individual basis and not only based on the thermal conditions. Recent studies have shown that occupants exposed to the same conditions could exhibit variations in environmental perception due to individual differences in comfort preferences and personality \citep{cheung2019analysis, livcina2018development}. Researchers have addressed this through the development of personal comfort models that predict individual thermal comfort responses rather than the average response of a larger population \citep{kim2018personal} through leveraging machine learning techniques and the Internet of Things (IoT) technologies. This approach demonstrates a very high prediction accuracy, well beyond that of the traditional models \citep{kim2018personal1}.

\subsubsection{Building systems' response to indoor environmental challenges}
In addition to the work in comfort models, a large amount of research has also been done to improve the systems that respond to comfort needs. Various mechanical systems technologies such as radiant systems and decentralized and personalized ventilation attempt to address comfort problems. These contemporary innovations focus on the ability of a building to adapt to its occupants by tracking them and modifying each person's immediate personal climate to meet their individual needs \citep{Brager2015EvolvingComfort}. 

Personalized control system approaches have limitations as the spatial resolution of most existing climate, lighting, and noise control technologies does not have enough flexibility and responsiveness. Even more innovative, decentralized and personalized comfort systems are unable to create the response and resolution needed to practically create individualized comfort zones for all occupants. Additionally, smaller and more decentralized systems create more maintenance tasks and complexity within the building systems \citep{VESELY2017223}. To meet these challenges of decentralization, a balance between personalizing spaces and maintaining the economies of scale that centralized systems provide could be achieved. Creating small zones of personal comfort for all occupants should be used sparingly in particular space use types, while most other spaces can be conditioned to meet the needs of a subgroup of people. 

\subsubsection{Collecting human comfort feedback in buildings}
Currently, researchers and building owners install a wide range of IoT devices that measure various environmental conditions such as light, noise, and particulates levels, in addition to the conventional temperature and humidity metrics. Although IoT sensors have become low cost and ubiquitous, the data from these devices are often not utilized to their full potential. Comfort models, even adaptive ones, only set thresholds to which sensor data points can be compared. The critical element in putting these data in context is subjective and physiological feedback from people who inhabit the environment. Collecting this type of data would empower more specialized and nuanced comfort models to be developed in a scalable way \citep{ltpaper}. However, the quantitative identification of individual differences in comfort preferences and personality of people remains a significant challenge in the field for researchers and practitioners \citep{WANG2018181}.

One way to tackle this situation is to use contemporary methods of personal feedback collection such as structured surveys or interviews either online or offline, in-person or remote, which would help increase the frequency and volume of building occupant feedback. However, such conventional methods have several shortcomings \citep{oecd}. One major drawback of these methods is the lack of scalability. It is difficult to collect large sample data sets due to the administrative, financial, and other operational overheads associated with these approaches. Furthermore, other factors such as lack of knowledge (respondents do not know the answer to a question, but answer it nonetheless), lack of motivation (respondents may not process questions fully) and failures in communication (survey questions may be unclear or misunderstood) result in an increased risk of biases and respondent heuristics in traditional survey responses \citep{bradburn2004asking}. As sensor adaptation in built environments continues to grow, new technologies and modern data capabilities allow researchers and practitioners to effectively capture dynamic human feedback effectively. However, this approach also presents significant challenges in the collection, analysis, processing, and visualization of large data sets from building occupants. There is a need for easy to use, scalable solutions that help identify and quantify occupant comfort preferences for operators as they move to ABW.

\begin{figure*}[ht!]
\centering
\includegraphics[scale=0.35]{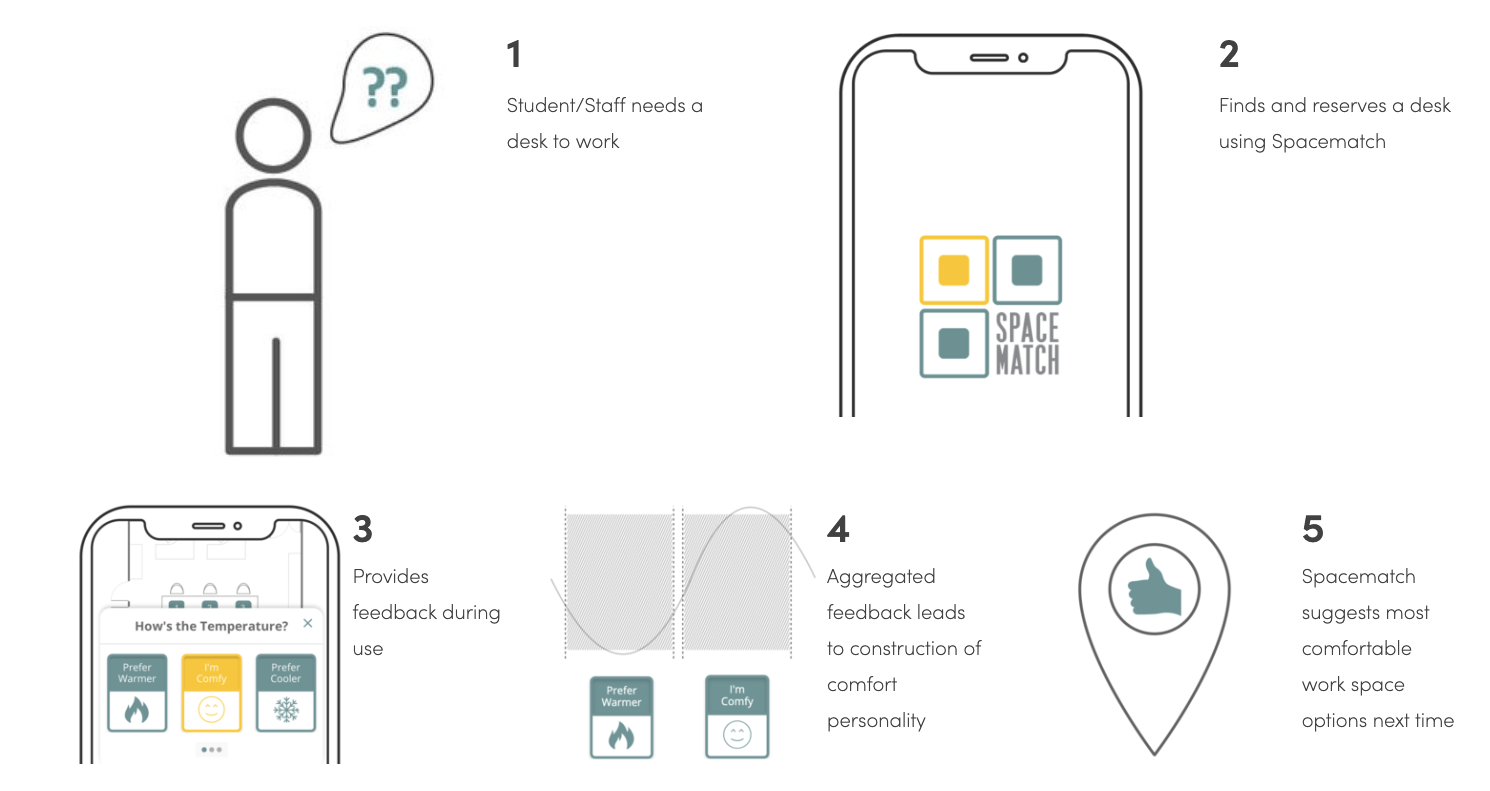}
\caption{Overview of the platform user flow - the goal is to give flexible workspace occupants the ability to find spaces that meet their needs (1 and 2), give feedback about their comfort to develop a comfort personality type (3), and eventually provide train a model (4) comfort suggestions for subsequent uses (5)} 
\label{userflow}
\end{figure*}

\begin{figure*}[ht!]
\centering
\includegraphics[scale=0.25]{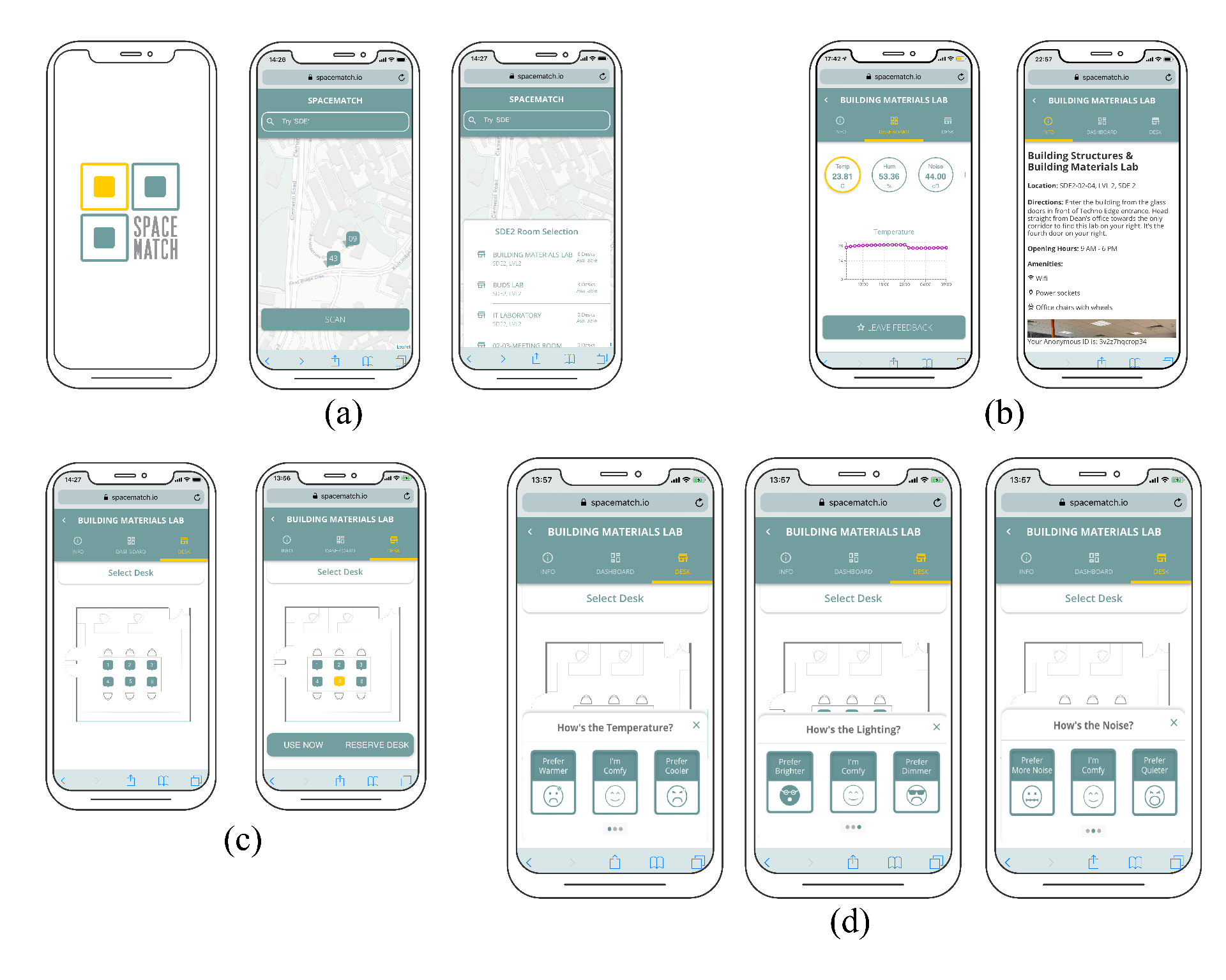}
\caption{Overview of the application: (a) Find flexible working zones around campus, (b) Dashboard and information screen for more details regarding each zone, (c) Choose between options for use desk now or reserve desk for later, (d) Provide feedback for temperature, light and noise variables} 
\label{app}
\end{figure*}

\subsection{Towards improving ABW by recommending the best location for an occupant based on their preferences}

This paper outlines a platform that improves indoor environmental satisfaction and ABW by allocating occupants to spaces that are the best match for their needs. We seek to test whether a longitudinally-intensive collection of indoor comfort data from individual occupants can be used to assign each person to a certain \emph{preference tendency type}. The goal is to use these preference types to match that person to a space that could best meet their needs. For example, consider a simple situation in which there are three occupants with indoor comfort preferences: Person A enjoys a dim, quiet, and moderately conditioned space, Person B is into a warmer and more active environment with a bit of noise and brightness, and Person C likes a cooler environment that is not dead silent. Perhaps these preferences are relatively consistent, or they could be dependent on the occupant's current activity or frame-of-mind. If these occupants' preferences are collected over time, the probability of nudging these users to spaces in offices that match these specific needs is higher. Also, there is the possibility of grouping people with similar preferences, which could be combined with systems control to create different zones with different types of comfort. Perhaps Person A, B, or C could be grouped with people who have comfort personality type \emph{A, B, or C} to improve both satisfaction and systems control. The goal of this study is to try to capture these tendencies using a web-based tool over a more extended period than a typical indoor comfort study.

\subsubsection{Novelty of proposed approach}
There have been several recent efforts with the focus of using technology to improve the ABW paradigm. One primary direction has been on the use of occupancy detection and prediction to characterize the use of ABW \citep{Rahaman2019-qd}. Additional work focuses on the use of human-sensor interaction to promote better decisions by occupants \citep{Arakawa2020-oq}. A previous study has used occupancy data to optimize the allocation of hot desk spaces using simulation and occupancy sensors \citep{Cooper2017AnData}. In the literature, there is a single case of designing a \emph{seat recommendation system} that has initial efforts towards matching people to spaces that would best match their needs \citep{Bae2014-vg}. Therefore, the presented platform and methodology is among the first examples of data collection, characterization, and efforts towards a space recommendation system. The presented scope of work combines the use of clustering occupants based on their preferences gathered in a longitudinally-intensive method and characterizing a comfort matching preference probability. These are techniques which have not been found in the literature.

\subsubsection{Organization of the paper}
This study addresses each of the previously mentioned challenges: collecting larger volumes of personalized data useful to occupants and operators, reducing the need for complex and problematic personalized comfort systems, and impacting ABW by improving spatial utilization and occupant comfort. This recommendation system's hypothesis is to test whether certain groups of occupants can be segmented according to their comfort preferences and whether this segmentation is realistic in the context of an actual building case study. This paper shares the development and testing of the platform in the context of a case study implementation. In Section \ref{section:methods}, we first illustrate an implementation with 25 research participants over a month in six flexible workspaces. Section \ref{Section:Results} showcases the results from this implementation and Section \ref{section:Discussion} discusses the interpretation of those results in the context of future work.

\section{Methodology}
\label{section:methods}

To test the user segmentation and space allocation hypothesis, a progressive web application platform was developed for implementation in the SDE4 and SDE2 building on the campus of the National University of Singapore. The first step was the development of the user flow of the platform. Figure \ref{userflow} illustrates the user in the case study who can find available open workspaces, reserve them for use, and give comfort feedback to help train a model to predict which location would be best for them based on their past comfort feedback. To use the platform, the occupants can choose to use the work desk right away or reserve for later use. During use, the mobile application enables occupants to quickly provide environmental comfort feedback for temperature, noise, and light variables, as shown in Figure \ref{app}. For flexible workspace operators, it facilitates merging personalized environmental comfort data with other data streams \emph{such as indoor environmental quality, occupancy, and energy use data among others} to optimize space, energy usage, and occupant comfort by grouping users with similar preferences.

\subsection{Implementation}

\begin{figure*}[ht!]
\centering
\includegraphics[scale=0.7]{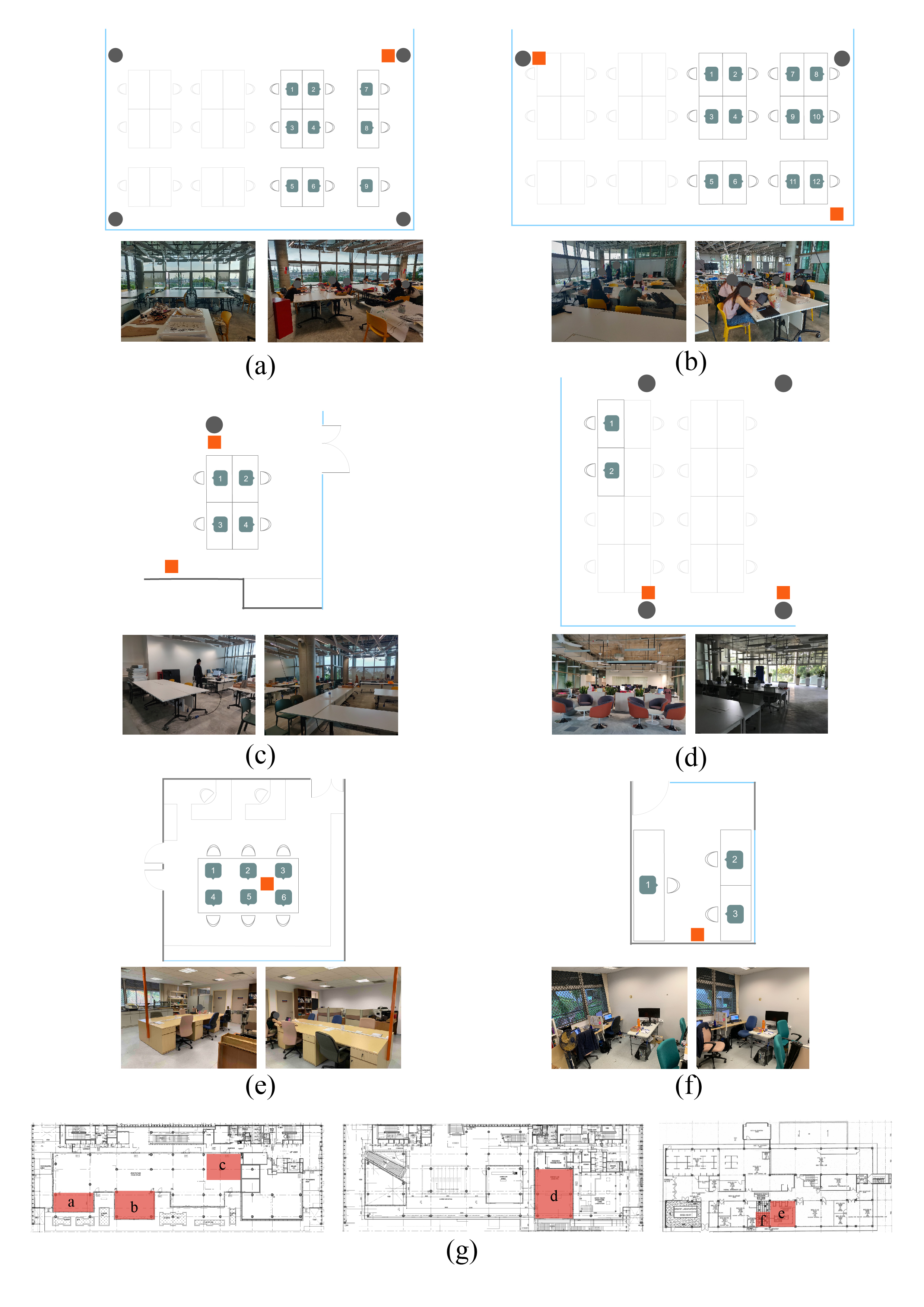}
\caption{The six flexible working zones for case study implementation. The orange square in each zone layout represents the location of the indoor environmental quality sensors: (a) Layout and photo from \emph{Zone 1}, (b) \emph{Zone 2}, (c) \emph{Zone 3}, (d) \emph{Zone 4}, (e) \emph{Zone 5}, (f) \emph{Zone 6}, (g) Selection of the six zones across three floors between two separate institutional buildings}
\label{pilot}
\end{figure*}

\begin{figure*}[ht!]
\centering
\includegraphics[scale=0.2]{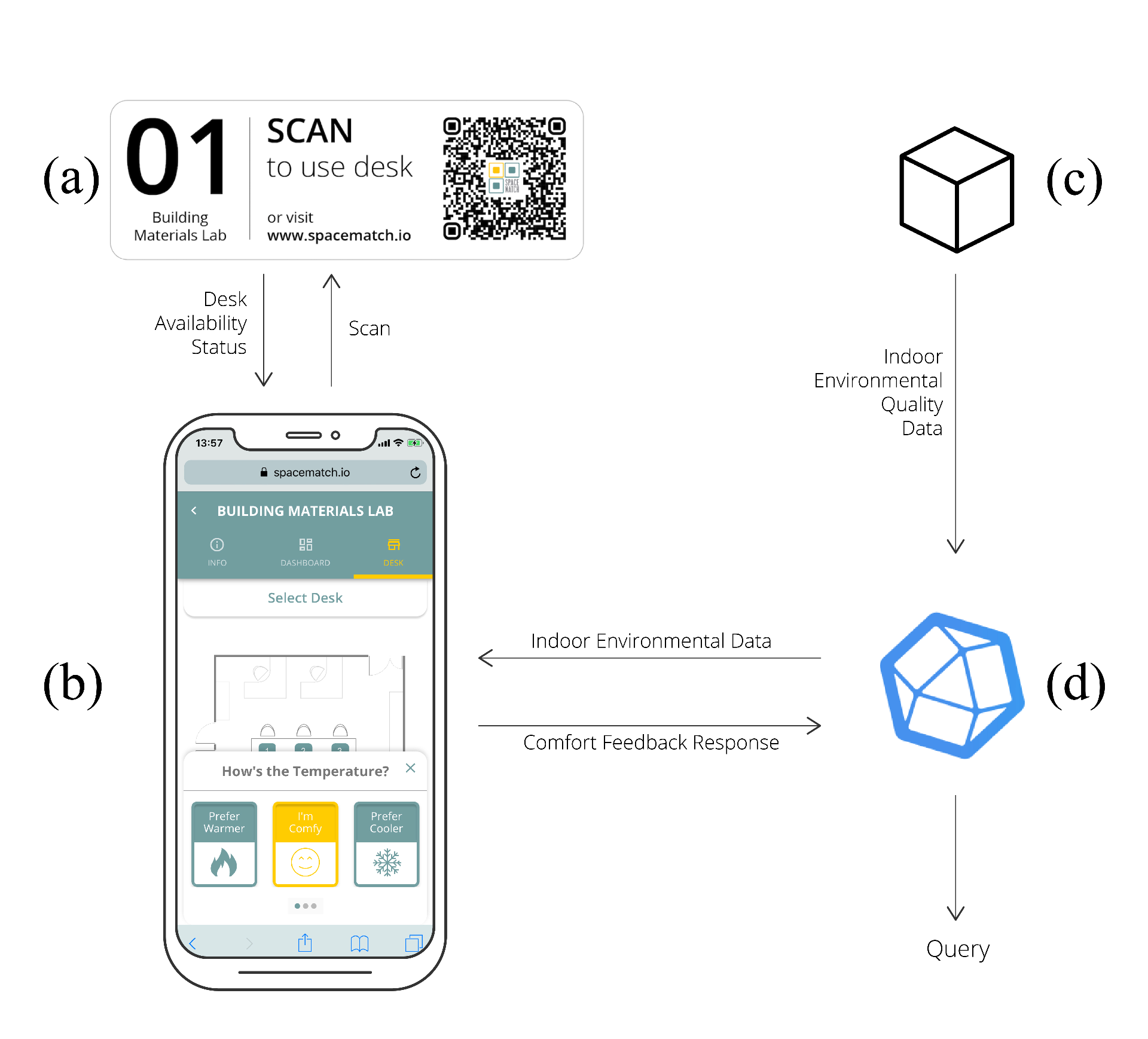}
\caption{Data exchange framework: (a) Desk QR code label, (b) Web-based mobile application, (c) IoT based indoor environmental quality sensors, (d) Time series data base} 
\label{framework}
\end{figure*}

\begin{figure*}
\centering
\includegraphics[scale=0.75]{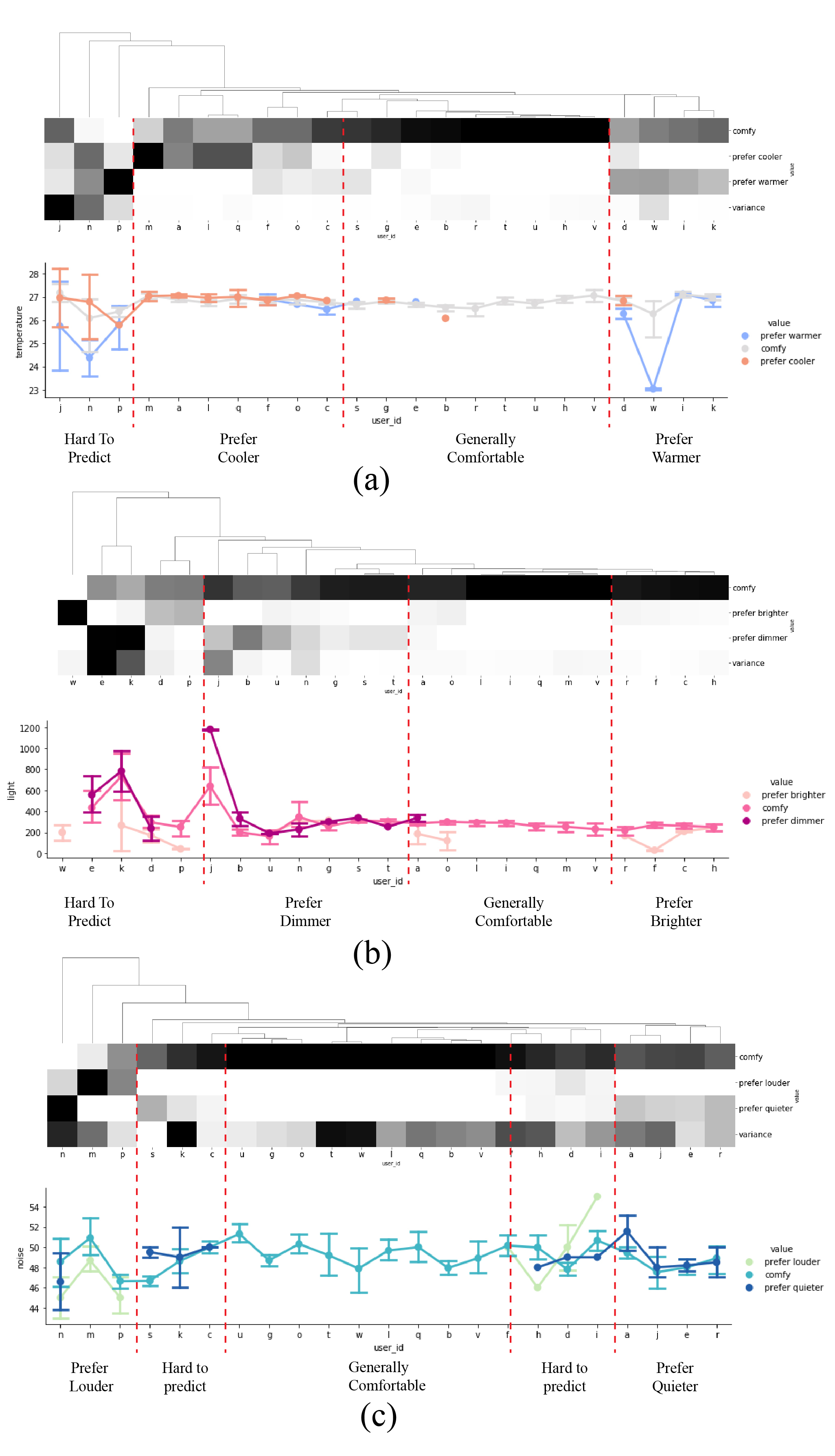}
\caption{User comfort preference clustering based on: (a) thermal comfort feedback (degrees Celcius), (b) noise comfort feedback (dB), (c) light comfort feedback (lux)} 
\label{clustering}
\end{figure*}

A case study was designed to test the platform in field conditions with 25 research participants over one month in April and May 2019. The participants were recruited from the current mix of students and staff, which were representative of regular users of two selected institutional buildings. An ethics review was submitted approved for the methodology of the study. A total of 36 desks in six different zones were identified for case study implementation, as shown in Figure \ref{pilot}. The zones were split between the two institutional buildings across three different levels. Each zone was strategically selected based on differences in location, floor level, number of desks, and zone accessibility. Further, differences in window-to-wall ratios, zone orientation, and proximity to the nearby vehicular road and public areas of the institution lead to varying light and noise levels between zones.

As shown in Figure \ref{pilot}, desks in each zone were arranged in the proximity of fixed indoor environmental quality sensors measuring seven attributes in real-time: temperature, humidity, noise, light, carbon dioxide, volatile organic compounds, and presence. Each zone offered flexible work desks in which each arrangement differed slightly. In some zones, desks were arranged to promote collaboration; they were aligned so participants could face each other while working. In others, the arrangement enabled solitary, personal work. Each desk was identified through a unique label containing the desk number, room name, and a QR code for connecting the desk to the progressive web application, as shown in Figure \ref{framework}.  During the study, participants were not assigned a specific desk and were encouraged to alternate their desk usage between different zones rather than stick to a single zone. The experimental instruction were designed help to provide generality and ecological validity to the collected data.

Each participant in the experiment used the interactive mobile application to reserve and use work desks, as shown in Figure \ref{app}. Participants could search for available work desks within the two institutional buildings. Once they chose the building, participants could progress to select the room and desk to use. They were provided with more information about the room and real-time indoor environmental quality through the \emph{Info} and \emph{Dashboard} features of the application, as shown in Figure \ref{app}b. The application provided options between using the desk right now or reserving it for later. To start using a desk, participants had to scan the desk QR code label using an in-built QR code scanner in the application. A minimum of a two-hour time slot for a work session was provided for each desk booking. Participants could also choose to extend their work sessions in two-hour multiples as per their requirements using the application. During use, the application prompted participants to provide environmental feedback for temperature, light, and noise levels through a three-point scale, as shown in Figure \ref{app}d. Prompts were configured such that the application nudged users to provide feedback at the start, finish, and once every half an hour of a typical two-hour work session. This miniature survey is a type of \emph{ecological momentary assessment}, a method for longitudinal data collection pioneered in medicine and psychology \citep{Moskowitz2006-sl} and recently adapted more for environmental perception \citep{Engelen2019-ti}.

At the start of the pilot study, a common onboarding session for all participants was organized. During this session, goals, objectives, and the methodology for the study were discussed, and the participants were onboarded to the interactive mobile application. Using the platform, the research team demonstrated to the participants on how to find and book a desk and to provide feedback during desk usage. Details such as participation schedule, timings, zone locations, and physical accessibility guidelines were also shared during this session. After this session, flexibility was provided for participants to use any of the six zones between 8 am to 6 pm daily for the month during the pilot study. Participants could also choose to participate in groups or individually based on their routine personal and work preferences to ensure that there is no disturbance to the typical flexible workspace environment. However, participants were encouraged to alternate their desk usage between different zones and times of the day to provide variety, generality, and ecological validity to the experimental findings. The test participants gave individual feedback in a range of 40-100 total feedback points per person. 

The data from the users and fixed sensors were aggregated using a cloud-based, time-series database, which served as a platform for data acquisition, storage, and error detection, as shown in Figure \ref{framework}. The combination of location-based user comfort feedback and fixed environmental sensor data allowed clustering analysis for personalized comfort profiles of users. These two data sources were merged through matching feedback location (spatially localized through desk QR code label), time of feedback collection (timestamp), and user ID (through an in-built anonymous authentication method).

\section{Results}
\label{Section:Results}
This section analyzes the environmental quality comfort preferences for temperature, light, noise values from 25 research participants in the pilot study. An unsupervised clustering technique is applied to segment participant comfort data, totaling 1,182 feedback points, into clusters based on similar behavior.  This study focuses on a participant's behavior based on their interaction with the system rather than on conventional variables in similar environmental preference studies, such as demographics, physiological, or environmental conditions. The emphasis is to apply an unsupervised clustering technique to the occupant data to segment the users who provide more than five feedback points into cohorts of similar behavior. This type of analysis focuses on the characterization of comfort preferences in ways specific to each occupant, but generalizable by grouping similar preference behavior. This effort captures each user's behavior in their interaction with the system instead of the demographic, physiological, or environmental conditions variables that are typically addressed in environmental preference studies.

\subsection{Discovering occupant personal comfort preference types}
To cluster user preferences, unsupervised learning techniques were used to group the participants into cohorts with similar feedback for temperature, light, and noise variables, as shown in Figure \ref{clustering}.  The analysis leads to identifying four distinct clusters based on differences in preferences for temperature, light, and noise levels across participants. As shown in Figure \ref{clustering}a, many participants are generally comfortable across zones. However, some participants preferred cooler or warmer environments. For participants in \emph{Hard to Predict} cluster, more extensive and more diverse data streams are needed to understand their preferences better in the future. For visual comfort or light values related preferences, the clustering is evenly spread between \emph{prefer dimmer} and \emph{generally comfortable} choices, as shown in Figure \ref{clustering}b. As can be observed, most participants would prefer a change in their light settings across zones. As shown in Figure \ref{clustering}c, most participants were aurally comfortable with a few preferring a change in noise levels. Understanding a user's past preferences and identifying the similarities and differences in preferences can be used to provide personalized spatial recommendations to individual users. 


\section{Discussion}
\label{section:Discussion}

\begin{figure*}[ht!]
\centering
\includegraphics[scale=0.65]{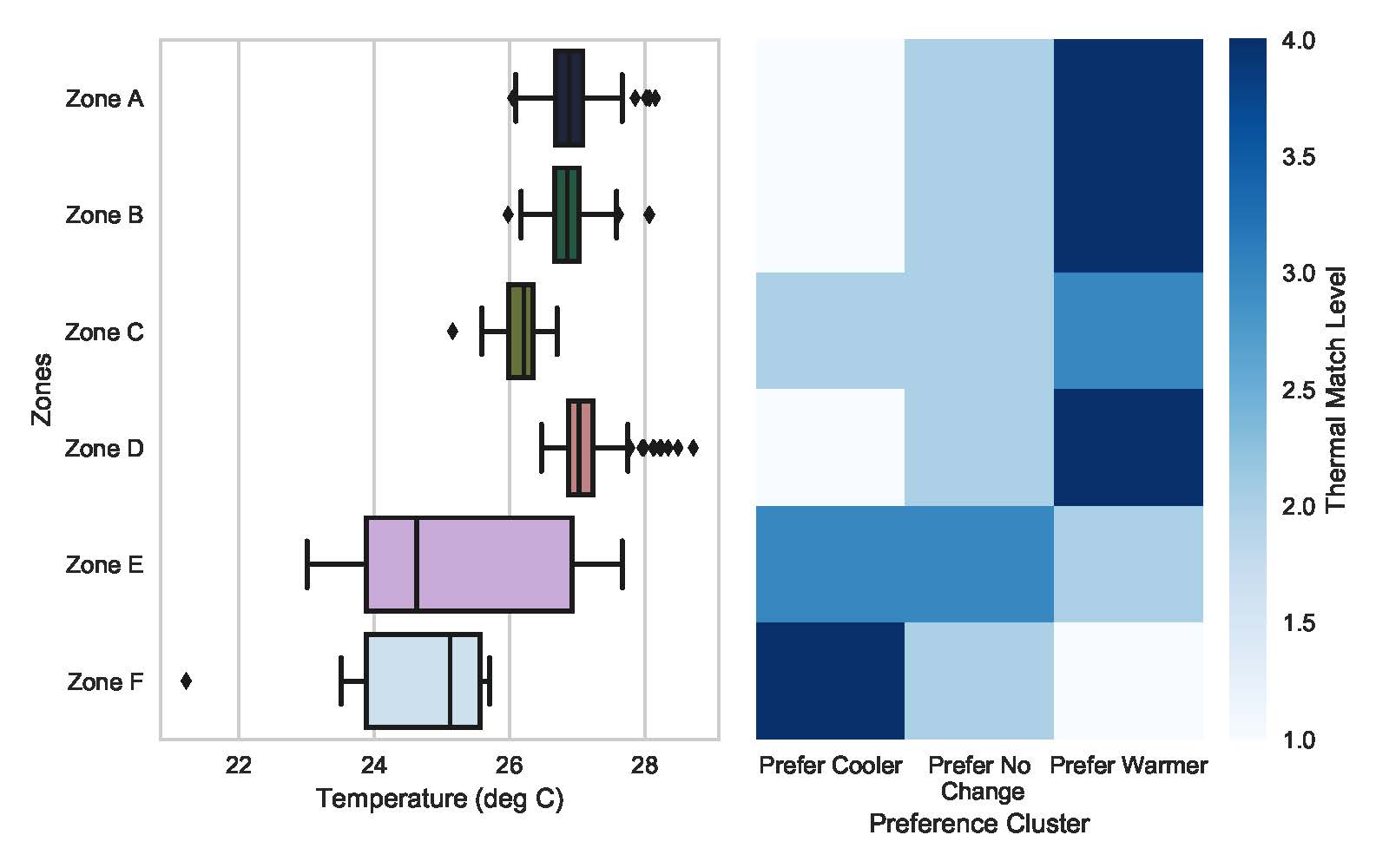}
\caption{Thermal comfort matching zone with preference type example. The box plots (left) illustrate the temperature ranges of each zone, while the heat map (right) shows the gradient of probability that each zone will be satisfactory for the comfort types segmented in Figure \ref{clustering}. Each of the zones has the potential to be a better or worse match for various comfort personality types - the higher the thermal match level, the more probability to meet the comfort preferences of each comfort personality type. This match level metric is the foundation for future work in the creation and testing of a recommendation system that automatically learns and suggests spaces to occupants.} 
\label{matching}
\end{figure*}

The implementation of the platform resulted in various insights related to the work towards a space recommendation system. Several lessons were learned from the related to deployment, the methodology used, the use of the data for comfort preference segmentation, and the foundation for automated means of space matching and allocation.

\subsection{Selecting a field-based experiment setup}
Thermal comfort research methodologies generally rely on two categories of implementation: laboratory-based methods (climate chambers) and field-based methods \citep{dedear}. However, while the internal design of a field-based study may not permit as rigorous statistical modeling and analysis as a carefully controlled climate chamber experiment, the field study serves a vital role in grounding the experimental findings in reality by its relevance to building occupants going about their regular daily routines. This methodology provides external validity to the experimental findings. It is also crucial for thermal comfort practitioners interested in understanding the role of the discipline of environmental psychology in building comfort.

Past studies have classified the laboratory-based and field-based methodologies as two fundamentally separate approaches; a deterministic \emph{engineering} approach versus a holistic person-environment system \emph{architectural} approach. The two approaches differ based on the disciplines which conduct them and their perception of the \emph{dynamic or static} relationship between occupant and buildings \citep{dedear}. However, models derived from a deterministic approach, work well only within limited conditions, usually centrally controlled air-conditioned spaces. That scenario can be compared to holistic person-environment systems models that take into consideration a more extensive range of conditions that building occupants may choose to make themselves comfortable, such as in naturally ventilated buildings \citep{dedear2, de1998developing}. Since one of the goals of this study is to understand the dynamic nature of occupant comfort in different environmental and spatial contexts - the research team chose a field-based experiment set up to provide higher ecological validity to the findings compared to a lab experiment \citep{andrade2018internal}.

\subsection{Using longitudinal data and a three-point preference scale}
New technologies have made collection, processing, and analysis of large and complex data more manageable. Using these capabilities, this study utilizes QR codes, a mobile application, and a time-series database infrastructure for management of the data life cycle. It enables the processing and assessment of a comparatively large comfort data set in a short time. Past studies in thermal comfort research have often referred to as five or seven-point scales. While valuable in some instances, recent work in this area has shown that user-friendly, simplified, and easy-to-use measures can be employed without the loss of predictive reliability and validity \citep{dolnicar2007user, krosnick2018questionnaire, dolnicar2011three}. For this study, the team used a three-point preference scale rather than the traditional seven-point thermal sensation scale, to limit subjectivity and make it easier for participants to frequently provide feedback in field conditions. This saved participant's effort and time in the field as well as helped channelize and organize data for the research team to work efficiently. In general, the aim of the comfort feedback prompts in the application (as shown in Figure \ref{app}d) was to seek answers to the following questions from participants regarding their perception of comfort: 1) Is their current condition comfortable or uncomfortable?; 2) Do they desire any change?; and 3) If so, would they prefer warmer or cooler?. From a psychological point of view, the first question relates to the cognitive thermal state and the other two to the preferred thermal state based on previous studies \citep{doi:10.1080/09613218.2016.1183185, parsons2014human}.         

\subsection{Identifying occupant comfort personality types}
Researchers increasingly adapt data-driven methodologies to address challenges of occupant satisfaction, environmental quality, and energy efficiency in buildings today \citep{Textmining, occupantcomplaints, IEQandsatisfaction}. This study uses data-driven methods to identify personalized comfort profiles of users - clustering users into \emph{types} based on similar environmental preferences, as shown in Section \ref{Section:Results}. Such results could be useful in multiple ways; for one, grouping users with similar environmental preferences could improve occupant comfort, space, and energy efficiency, as shown by other studies \citep{KAMARULZAMAN2011262}. Next, this method could also enhance the feedback given to designers and operators about future building design features and operating strategies to improve flexible workplace occupant satisfaction and performance \citep{KWON2019356}. In parallel, it is also easy to see how the same methodology could be used to distinguish spaces based on occupant comfort feedback data and IoT data - to derive comfort profile types of spaces.

\subsection{Towards a space recommendation engine}
It is easy to see how the results from this study could be used to understand, and even predict, patterns and anomalies in an occupant's environmental preferences in flexible workspaces over time. Taking this a step further, learning from past comfort preferences of occupants could be used to \emph{match} them to spaces with suitable environmental profiles with acceptable temperature, light, and noise levels on average. This process can be done in real-time using IoT data to test whether this leads to an increase in occupant satisfaction or performance in flexible workplaces compared to a baseline scenario. Such methods of suggesting or matching based on past preferences have been widely used in other industries such as media and social networking \citep{Amatriain2015, Recommender, Resnick}, but they are still a new concept for the built environment.

Figure \ref{matching} illustrates an example of this potential matching paradigm as applied to thermal comfort. This figure shows the distributions of dry bulb temperature for each of the Zones from the case study as well as a heat map illustrating a subjectively-selected \emph{Thermal Match Level} based on the feedback illustrated in Figure \ref{clustering}. These results show how the segmentation created by the occupant data feedback histories can be used to match them to spaces according to the match level. There is also the potential to interact with the building control systems to change the conditions of the spaces to create more or less of a \emph{space type} based on the changing needs of building occupants. This paper illustrates that the collection of data and segmentation of users is possible using the type of feedback data collected from the test participants. Future work will investigate how this process can be automated and metrics developed to show the success of such matching in terms of reducing energy consumption and improving thermal comfort.

\section{Conclusion}
This paper describes the field-based implementation of a space allocation platform in six flexible working zones for occupant comfort data collection. Over a month, 25 participants provided 1,182 environmental momentary assessment surveys of their thermal, visual, and aural comfort. This comprehensive data set provides exciting opportunities for interpreting and learning about occupant comfort behavior in built environments through data-driven methods. By demonstrating how data can be utilized to group occupants into comfort profile types, this study can act as a potential stepping stone to related research areas such as comfort profiling of spaces, occupant behavior analysis, and correlation identification between various spatiotemporal variables in buildings. 

\subsection{Limitations}
This analysis has covered the deployment and collection of data from users of a matching-based flexible platform. However, it has stopped short of testing the ability to give the recommendations and the reactions of users in the face of these suggestions. An additional limitation for this study is that the sample size of participants is not large enough to make more generalizable characterizations of the comfort types and the vast range of behavior that occupants could exhibit. Also, the number and type of physical measurements in the spaces were not exhaustive as phenomena such as radiant and space effects were not measured. The next phase of the project is a \emph{spatial recommendation engine} seeks to test the feature to suggest spaces to people in order for them to find available working spaces that match their immediate needs. This deployment could be framed in the same way that common platforms help people find a place to stay or find a ride. This platform design would then test technologies such as desk recommendation (based on time duration, number of desks, noise levels and desk availability), and integration with occupancy, intelligent power plugs, and building management systems. Future deployments of the platform will focus on the data collection from a larger sample size, enabling a much more generalizable characterization of comfort groups and measuring more physical parameters.  

\section*{Conflict of Interest Statement}
The authors declare that the research was conducted in the absence of any current commercial or financial relationships that could be construed as a potential conflict of interest.

\section*{Author Contributions}
TS: platform design, infrastructure development, experimental design, implementation and lead author of the paper; PJ: conceptualization, author of the paper; CM: experimental design, data analysis, funding, project leadership, the corresponding author of the paper.

\section*{Funding}
The Ministry of Education (MOE) of Singapore (R296000181133) and the National University of Singapore (R296000158646) provided support for the development and implementation of this research. 

\section*{Acknowledgments}
The authors would like to acknowledge Prageeth Jayathissa and Matias Quintana for their assistance in data collection and processing, the NUS Department of Architecture and Building for the use of their spaces, and Naor Biton's efforts in the development of the platform.

\section*{Data Availability Statement}
Segments of the raw data and analysis code used for this study will be available in an open-access Github repository that includes further documentation\footnote{\url{https://github.com/buds-lab/spacematch-paper}}. 

\bibliographystyle{model1-num-names}
\bibliography{spacematch}

\end{document}